\documentclass[conference]{IEEEtran}
\IEEEoverridecommandlockouts
\usepackage{cite}
\usepackage{graphicx}
\usepackage{textcomp}
\usepackage{rotating}
\usepackage[inkscapelatex=false]{}
\usepackage{enumitem}
\usepackage{placeins}  % For \FloatBarrier
\usepackage{tikz}

\newcommand\copyrighttext{%
	\footnotesize \textcopyright 2024 IEEE. Personal use of this material is permitted.
	Permission from IEEE must be obtained for all other uses, in any current or future
	media, including reprinting/republishing this material for advertising or promotional
	purposes, creating new collective works, for resale or redistribution to servers or
	lists, or reuse of any copyrighted component of this work in other works.
	}
\newcommand\copyrightnotice{%
	\begin{tikzpicture}[remember picture,overlay]
		\node[anchor=south,yshift=10pt] at (current page.south) {\fbox{\parbox{\dimexpr\textwidth-\fboxsep-\fboxrule\relax}{\copyrighttext}}};
	\end{tikzpicture}%
}

\def\BibTeX{{\rm B\kern-.05em{\sc i\kern-.025em b}\kern-.08em
    T\kern-.1667em\lower.7ex\hbox{E}\kern-.125emX}}

\begin{document}

\title{Cross-course Process Mining of Student Clickstream Data - Aggregation and Group Comparison\\
\thanks{The study was financially supported by the Vienna Municipal Department for Economic Affairs, Labour and Statistics as part of the project “Teaching \& Learning Analytics for data-based optimization of teaching and learning processes in courses with blended learning”, which was granted in the 32nd call for Quality Assurance of Teaching at the Viennese Universities of Applied Sciences. }
}

\author{\IEEEauthorblockN{1\textsuperscript{st} Tobias Hildebrandt}
\IEEEauthorblockA{\textit{Department of Computer Science} \\
\textit{University of Applied Sciences Technikum Wien}\\
Vienna, Austria \\
tobias.hildebrandt@technikum-wien.at}
\and
\IEEEauthorblockN{2\textsuperscript{nd} Lars Mehnen}
\IEEEauthorblockA{\textit{Department of Computer Science} \\
	\textit{University of Applied Sciences Technikum Wien}\\
Vienna, Austria \\
lars.mehnen@technikum-wien.at}
}

\maketitle
\copyrightnotice

\begin{abstract}
This paper introduces novel methods for preparing and analyzing student interaction data extracted from course management systems like Moodle to facilitate process mining, like the creation of graphs that show the process flow. Such graphs can get very complex as Moodle courses can contain hundreds of different activities, which makes it difficult to compare the paths of different student cohorts. Moreover, existing research often confines its focus to individual courses, overlooking potential patterns that may transcend course boundaries.

Our research addresses these challenges by implementing an automated dataflow that directly queries data from the Moodle database via SQL, offering the flexibility of filtering on individual courses if needed. In addition to analyzing individual Moodle activities, we explore patterns at an aggregated course section level. Furthermore, we present a method for standardizing section labels across courses, facilitating cross-course analysis to uncover broader usage patterns.

Our findings reveal, among other insights, that higher-performing students demonstrate a propensity to engage more frequently with available activities and exhibit more dynamic movement between objects. While these patterns are discernible when analyzing individual course activity-events, they become more pronounced when aggregated to the section level and analyzed across multiple courses.

\end{abstract}


\section{Introduction}
Learning Management Systems (LMS) like Moodle are increasingly utilized in university courses, offering a plethora of features beyond simple content delivery. These include functionalities such as assignment uploads, automated self-checks with grading and feedback capabilities, voting systems, and forums. As Moodle courses become more multifaceted, incorporating functionalities previously scattered across different systems, instructors are keen to ascertain whether students are engaging with courses as intended and whether course designs can better cater to student needs.

This research deals with understanding the process flow within courses – the sequence in which students navigate course elements. Visualizing and analyzing such flows benefits from the application of methods from process analytics and process mining, which involve the analysis of activities and their sequential order from real-life event logs\cite{van_der_aalst_process_2011}. In this context, Moodle activities can be conceptualized as process activities within the context of an event log, while a student's journey through a Moodle course can be viewed as a process instance.
While a substantial body of research exists that bases on this paradigm, often focusing on detailed analyses of process flows within specific courses, our paper adopts a different approach. We delve into the retrieval and analysis of aggregated data at both the course-section level as well as the potential for course-spanning process mining. By shifting the focus from individual activities to course sections, we aim to provide a broader understanding of student interaction patterns. Additionally, our exploration of course-spanning process mining opens avenues for uncovering insights that transcend individual courses, offering a more comprehensive view of student engagement and learning behaviors.
The analysis is based on a substantial sample from the Moodle database of our institution. The base sample before data preparation includes around 3,4 million events from 5,993 unique students that have received 20,131 grades in 2,096 different courses.
After an overview of relevant related work, we presents the applied methods of data engineering and process mining. The subsequent section presents the results in terms of data engineering, process analytics and course-spanning path comparison. The paper continues with the discussion of the results and ends with the conclusion.

\section{Related Work}
Studies such as \cite{ademi_prediction_2019} show, that there are positive correlations between students' interactions with LMS courses and their grades, with predictors such as course visits, quiz attemts, participation in assignments and submissions.
However, these analysis do not investigate whether the sequence in which these activities are performed might also differ depending on the course performance.

A comprehensive step into the direction of making such event logs usable for the purpose of educational process mining has been proposed by Bolt et al \cite{bolt_business_2017}. The authors present an automatable workflow using the RapidMiner toolkit which extracts event data and generates reports based on the OLAP principle (On-line Analytical Processing), a concept primarily used in the domain of Business Intelligence.
In their proposed system, event-based statistics as well as students process flows can be generated by applying OLAP operations such as filtering, roll-up or drill-down across data dimensions such as time (e.g. by selected a time frame) or courses (by selected one or more specific courses for analysis), but not on the granularity of the events itself (e.g. switching between individual course activities or the course sections they belong to).
Specific challenges that arise when creating process maps based on cross-course data instead of from individual courses were not investigated.

There exists a wide range of research that deals with analyzing LMS data to investigate students' learning paths, like \cite{bakar_process_2019}. Others have taken a more comprehensive approach by combining clickstream data with learners' self-report questionnaires, as exemplified by  \cite{maldonado-mahauad_mining_2018}. The authors categorized student logs into two groups: those who completed the course and those who did not. They then employed techniques such as clustering to mine various interaction patterns, which they matched with different types of learners. A similar methodology was adopted in\cite{mukala_learning_2015} and \cite{real_educational_2020}, who conducted process mining on individual events for student groups based on their course performance and compared their process flows. The latter also introduced initial steps toward considering event hierarchies, although they focused on component types (such as tasks and quizzes) instead of course sections.

\cite{moreno_process_2021} presented the differences in the learning paths between constructing the process map from individual LMS activities on the one hand, and grouping those individual activities to course sections on the other hand. It is shown that grouping based on section results in a more structured process map. The authors did not investigate how those process maps differ among groups with different course performances.
Others have tried to mine different learners types directly from the LMS clickstream data \cite{crosslin_understanding_2021}.  The authors found three types of learners and present a different process graph for each.

Currently, there are limited tools available that facilitate comparisons between user groups by visualizing differences in a single process map. One tool that enables this is the Process Comparator \cite{bolt_visual_2016}, a plugin for the open source Process Mining Suite ProM. This plugin enables the visualization of statistically significant differences between two traces of the same process model, both in terms of the frequency of certain activities and how often direct paths between activities are taken.

Researchers such as \cite{thiyagarajan_process_2021} have utilized this plugin to compare the paths of different student groups based on their course performance. Their study did not focus on
the comparison regarding how often the different student groups have used the different Moodle activities or whether one group used specific paths significantly more often than the other, but rather on the differences concerning elapsed time between activities.

To the best of our knowledge, there is a gap in research that

a) compares students' paths not only based on direct clickstream data but also on aggregated section-level data

b) compares process flows in a cross-course manner, enabling insights that transcend individual courses and provide a broader understanding of student engagement and learning behaviors.

We are interested in the following questions:
\begin{enumerate}[label=\Roman*.]
	\item How can LMS clickstream data be prepared to enable cross-course process analysis? 
	\item What are the distinct characteristics of LMS paths among high-performing students compared to their counterparts with lower academic achievements?
	\subitem Do the observed patterns in LMS-path behavior vary when individual events are aggregated into course sections, compared to analyzing them at the individual event level?	
	\item Are there discernible cross-course patterns in student learning paths?
	\subitem  Do these patterns vary based on academic performance?
\end{enumerate}

\section{Methods}
This section describes the applied methods of data engineering and process mining.
\subsection{Data Engineering}
\subsubsection{Data Retrieval}
The data were extracted from the anonymized Moodle database, encompassing information from all courses during the winter semester of 2022. Two distinct SQL statements were utilized to extract the necessary data, resulting in the generation of two files:

\begin{enumerate}[label=\Roman*.]
	\item the event log with the following columns (see table \ref{tab:event_log})
	\item  a record with the final score (between 0 and 100) for each student for each course
\end{enumerate}

\begin{table*}[htbp]
	\centering
	\caption{Event Log columns}
	\label{tab:event_log}
	%\scriptsize
    \begin{tabular*} {\textwidth}{lllllll}%{@{}llllll@{}}
	 	\hline
		Timestamp & Course Name & CourseID & Event & Section & UserID \\
	 	\hline
		% Add your data rows here
		2022-08-30 17:25:20.000 +0200 & Mathematics for Computer Scientists & 1 & Download  2 & Section A & user1 \\
		2022-09-05 19:26:03.000 +0200 & Physics & 2 & Exam 1 & Section B & user2 \\
		2022-09-09 13:27:55.000 +0200 & Computer Science Introduction & 3 & Template & Section C & user3 \\
		% Add more rows if needed
		\hline
	\end{tabular*}
\end{table*}

The SQL query for the event log was designed to filter the data, exclusively including events associated with student interactions. Additionally, it excluded courses that had been deleted during the specified timeframe and those with fewer than 100 events. This filtering mechanism aimed to exclude courses potentially created for testing purposes or those not reliant on Moodle for instruction.

Similarly, the SQL query for the scores was filtered to encompass final grades falling within the range of 0 to 100.
Both sets of results were exported into separate CSV files. On average, the event logs contain around 1,600 events per course and approximately 160 events per student within a course.

It is important to note that only a subset of these logs was utilized for analysis. Many records included interactions by users who had been deleted in the meantime or who for other reasons never received a score. Additionally, numerous courses lacked scores altogether, as grading was managed externally or the courses primarily served informational purposes for students and employees. For instance, Moodle is frequently utilized internally for employee training and organizational functions.

After filtering out events that could not be connected to a score, approximately 1.8 million events remained, involving 4,361 unique students who received 18,406 grades across 747 courses. This translates to an average of approximately 2,400 events per course and around 100 events per student within a course.
\subsubsection{Data transformation}
The general process of preparing the data for analysis can be summarized using the following steps:
\begin{enumerate}[label=\Roman*.]
	\item Read in and merge CSV files
	\item 
	\begin{enumerate}
	\item For individual courses: filter for course name 
	\item For course-spanning analysis: homogenize section names and create new column consisting of courseId and studendId
	\end{enumerate}
	\item optional: remove self-loops
	\item Split rows based on performance into higher-performing and lower-performing half of students
	\item Export tables to XES files
	
\end{enumerate}

\begin{figure*}[htbp]
	\centerline{\includegraphics[width=0.9\textwidth]{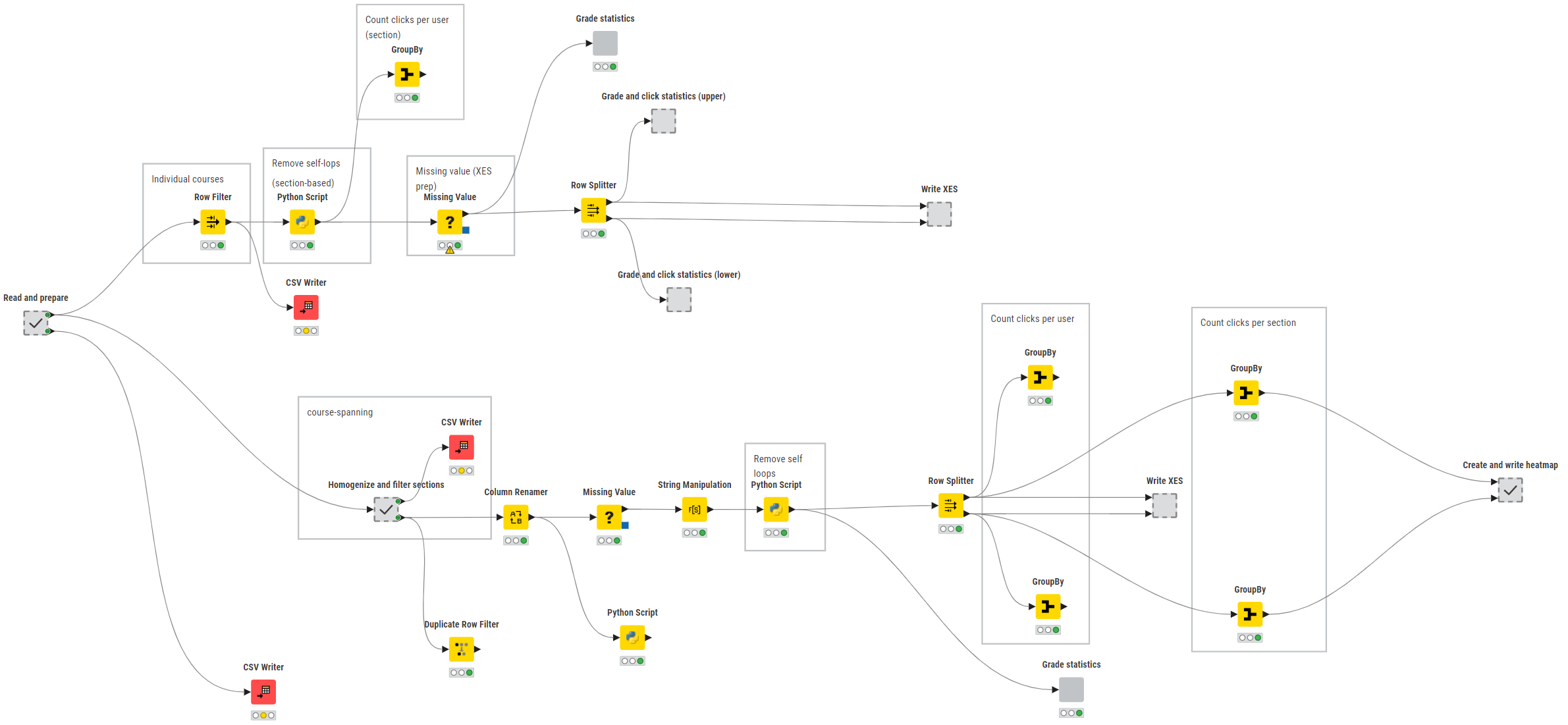}}
	\caption{Data Preparation Process.}
	\label{fig:KNIME1}
\end{figure*}

The data processing was conducted using the open-source tool KNIME, in conjunction with the PM4KNIME extension. The workflow involved merging files, applying additional filtering, preprocessing the data, generating statistics, grouping data, and ultimately exporting XES (eXtensible Event Stream) files for use in process mining tools. The exported XES files were organized into two groups: one containing event logs for the lower-performing half of students and the other for the higher-performing half. These files were exported for individual courses, each varying in levels of detail, as well as for cross-course analysis. A screenshot illustrating the entire process is provided in Figure \ref{fig:KNIME1}.

The detailed steps were the following:
\begin{enumerate}[label=\Roman*.]
 
\item The two CSV files containing the event log and student score records were read in (exported before from the Moodle database using SQL commands).
\item The tables were joined based on courseID and studentID. 
%\item If desired, the output can already be exported into a CSV file for analysis in tools like Fluxicon Disco
\item In order to generate a valid XES file, the timestring had to be converted to the data type \textit{date \& time}.
\item 
\begin{enumerate}
	\item For the course-based analysis, a filter node was inserted to enable to select only log entries that belong the course of interest.

\item For cross-course analysis, a table with a dictionary was created to summarize similar sections using regular expressions (see Table \ref{tab:KNIME2}). This is possible because many section names throughout different courses use a similiar naming structure - the sections designed for self study often either contain the terms \textit{self study} or \textit{self-study} (or the corresponding German word) followed by a blank and either a number (like \textit{self study 1}) or an alphabetic numbering system (like \textit{self study a}). Such section names were standardized to "self study "+number. Similarly, the sections designated for in-presence units were often named using the pattern "class" + number/letter (or the corresponding German word). Those section names were standardized to "class "+number.
\end{enumerate}

\begin{table}[hbp]
	\centering
	\begin{tabular}{|p{0.7\columnwidth}|p{0.2\columnwidth}|}
		\hline
		\textbf{Input} & \textbf{Output} \\ \hline
	(.*)(Class) (\d)(.*) & class \$3 \\ \hline
	(.*)(self-study) (\d)(.*) & self study \$3 \\ \hline
	(.*)(self study) (\d)(.*) & self study \$3 \\ \hline
	(.*)(eigenstudium) (\d)(.*) & self study \$3 \\ \hline
	(.*)(präsenz) (\d)(.*) & class \$3 \\ \hline
	(.*)(eigenstudium\textbar self study\textbar self-study\textbar eigenstudium) (a)(.*) & self study 1 \\ \hline
	(.*)(eigenstudium\textbar self study\textbar self-study\textbar eigenstudium) (b)(.*) & self study 2 \\ \hline
	(.*)(eigenstudium\textbar self study\textbar self-study\textbar eigenstudium) (c)(.*) & self study 3 \\ \hline
	(.*)(eigenstudium\textbar self study\textbar self-study\textbar eigenstudium) (d)(.*) & self study 4 \\ \hline
	(.*)(eigenstudium\textbar self study\textbar self-study\textbar eigenstudium) (e)(.*) & self study 5 \\ \hline
	(.*)(eigenstudium\textbar self study\textbar self-study\textbar eigenstudium) (f)(.*) & self study 6 \\ \hline
	(.*)(eigenstudium\textbar self study\textbar self-study\textbar eigenstudium) (g)(.*) & self study 7 \\ \hline
	(.*)(eigenstudium\textbar self study\textbar self-study\textbar eigenstudium) (h)(.*) & self study 8 \\ \hline
	(.*)(eigenstudium\textbar self study\textbar self-study\textbar eigenstudium) (i)(.*) & self study 9 \\ \hline
	\end{tabular}
	\caption{Section name harmonization for sections 1 to 9}
	\label{tab:KNIME2}
\end{table}

%\begin{figure}[htbp]
%	\centerline{\includegraphics[width=\columnwidth]{}}
%	\caption{Harmonization of section labels for cross-course analysis.}
%	\label{fig:KNIME2}
%\end{figure}

\subitem The \textit{String Replacer (dictionary)}-node was used to replace section titles based on the dictionary. Table \ref{tab:section_replacements} shows how many different section names were replaced with the respective aggregation titles. For example, 237 different section titles were replaced with "class 1".

\begin{table}[hbp]
	\centering
		\caption{Number of replaced section titles}
	\label{tab:section_replacements}
 	  \begin{tabular}{|p{0.5\columnwidth}|p{0.4\columnwidth}|}
		\hline
		Aggregated section title & Number of replaced titles \\
		\hline
		class 1 & 237 \\
		class 2 & 105 \\
		class 3 & 98 \\
		class 4 & 90 \\
		class 5 & 74 \\
		class 6 & 70 \\
		class 7 & 61 \\
		class 8 & 48 \\
		class 9 & 28 \\
		self study 1 & 114 \\
		self study 2 & 112 \\
		self study 3 & 97 \\
		self study 4 & 82 \\
		self study 5 & 75 \\
		self study 6 & 60 \\
		self study 7 & 56 \\
		self study 8 & 39 \\
		self study 9 & 35 \\
		\hline
	\end{tabular}

\end{table}

\subitem A filter was applied to include only section names that confine to the new schema.
After standardizing section names, there were 393 courses remaining.

\item For cross-course analysis, rows with missing values in event/section column were removed, ensuring compatibility with PM4KNIME.
\item A \textit{Row Splitter} for grouping students based on their performance into group A and group B was inserted.
\begin{enumerate}
	\item For individual courses,
students were split into two groups (Group A and Group B) based on their score in relation to the median score of the respective course. 
\item For cross-course analysis,
a new column was created containing both courseID and studentID.
The overall median score was calculated using a unique combination of courseID and studentID.
Students were split into two groups (Group A and Group B) based on the median score of the combined data.
It is not sufficient to calculate the median of all students, since one student can participate in more than one course, and each course potentially shows a different usage behavior. Instead, a new column containing both courseID and studentID has been created, for which the median was calculated. Thus, for each individual student that has participated in a specific course, a new caseid column has been created, that consists of courseID+studentID.
\end{enumerate}  
\item	For both groups the \textit{Table to Event Log} node was used.
\begin{enumerate}
\item	For individual courses, the anonymized studentID was selected as \textit{case} column. For the \textit{eventclass} column, alternately the activity name (individual events) or section (for section-based grouping) were used, depending on the desired analysis.
\item	For course-spanning analysis, the combination of courseID and studentID was used as \textit{eventclass}, whereas for activity the recently standardized section name was used. Individual events were not used as they will be hard to impossible to standardize between courses.
\end{enumerate}

\item Finally the XES Writer was used to write an output that can be read in by ProM.
\end{enumerate}

\subsection{Process Mining}
For the initial display of process maps and to generate an overview of the data, the process analytics tool Fluxicon Disco has been used. For the concrete analysis of statistical differences between the student groups, ProM in conjunction with the aforementioned Plugin Process Comparator has been used. The default settings for the transition system have been used, for the comparison settings we have opted to compare annotations using the process metric occurrence frequency. Both states and transitions have been included, for the alpha significance level the default value of 5\% has been used.
For reasons of variety, for the comparisons within individual courses, two courses have been selected. The bachelor course Procedural Languages is used, as the course had a larger number of students (210) and was taught in several groups. The master course Data Warehouse \& Business Intelligence is an example for a course with a smaller student cohort, as it comprised 34 students divided into 2 groups.

\section{Results}
\subsection{Data Engineering}
\subsubsection{Aggregation}
\begin{figure}[htbp]
	\centerline{\includegraphics[width=\columnwidth]{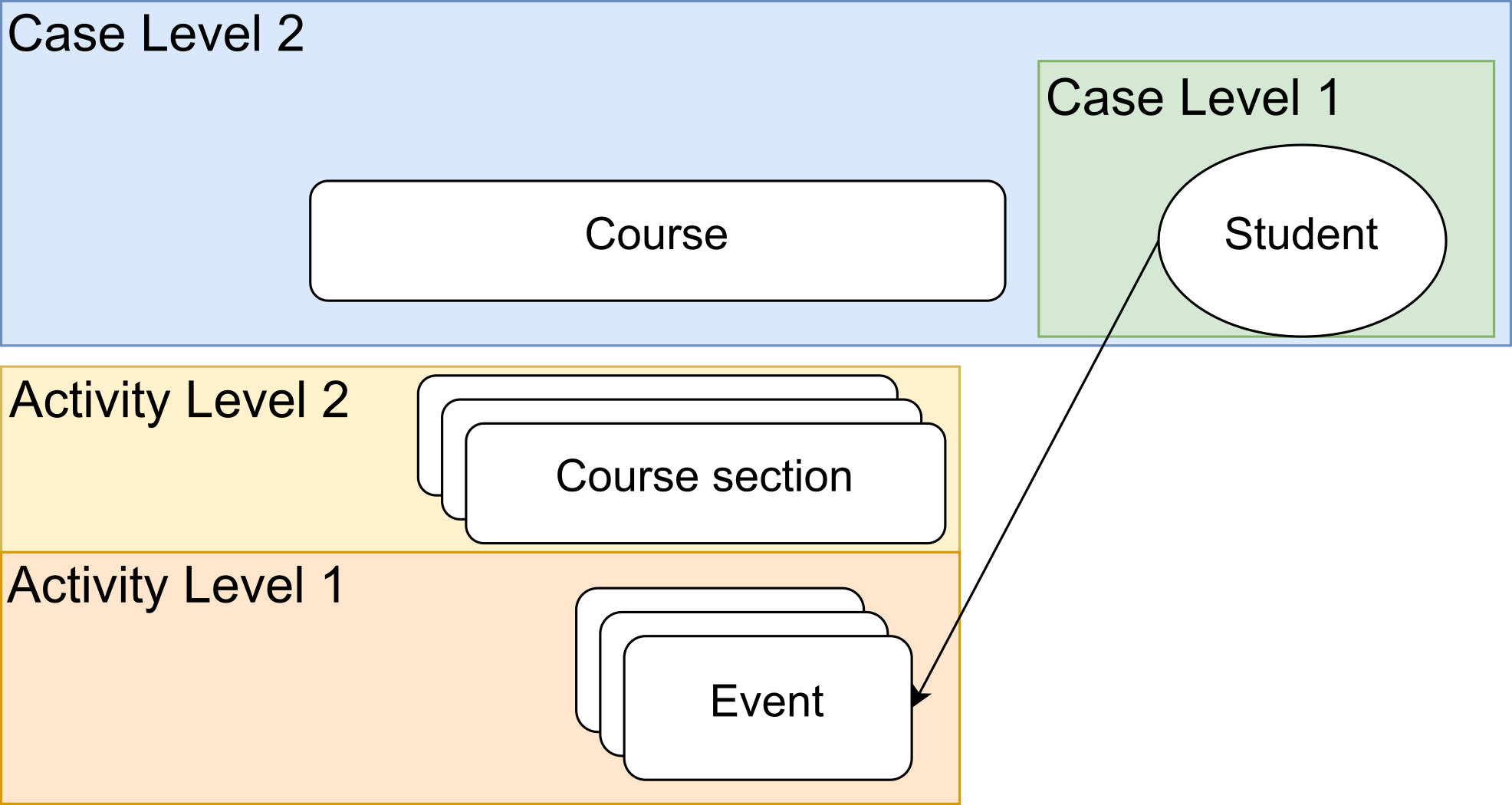}}
	\caption{Mapping from log data fields to XES parameters.}
	\label{fig:mapping}
\end{figure}

As previously described, various pairs of event logs were created, with each pair containing data for Group A (comprising the better-performing half of students) and Group B (comprising the other half). These three pairs were:
\begin{enumerate}[label=\Roman*.]
	\item course-specific; individual Moodle activities as XES-activity
	\item course-specific; section names as XES-activity
	\item course-crossing; homogenized section names as XES activity.
\end{enumerate}
The summary of the different mappings of event log data dimensions to the XES concepts of case and activity in the different levels of aggregation is illustrated in Figure \ref{fig:mapping}.

\subsubsection{Self-loops}
For both, intra-course and inter-course analysis, it may be interesting for certain statistics to ignore self-loops within a section. For example, if a student consumes several Moodle activities of the same section in a row, before proceeding to another section, those events are called \textit{self-loops}. For an analysis regarding how often students jump between sections of a course, it may make sense to remove such self loops. This has been achieved by creating an optional Python-node, which removes events in cases where the previous event by the same student in the same course did already belong to the same section.

\subsection{Process Analytics}
\subsubsection{Low-level-event-based analysis}
In the course \textbf{Procedural Languages}, the group was split based on the median score threshold of 70.6 out of 100.
In the lower-performing group, there were 43,129 events recorded. In contrast, the higher-performing group had 45,824 events recorded.

\begin{figure}[htbp]
	\centerline{\includegraphics[width=\columnwidth]{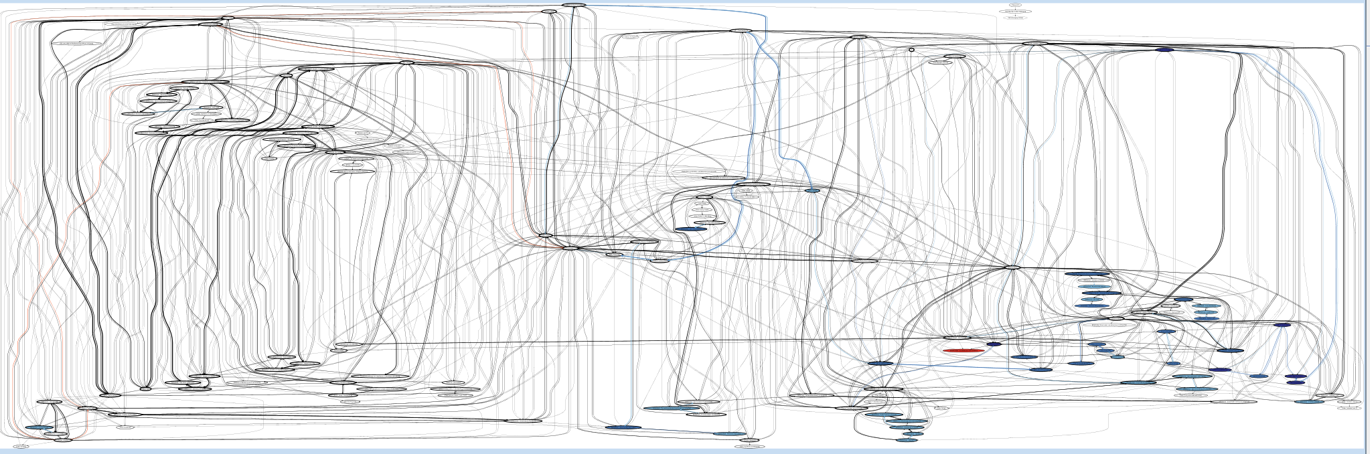}}
	\caption{Procedural Languages without filtering.}
	\label{fig:proc1}
\end{figure}

\begin{figure}[htbp]
	\centerline{\includegraphics[width=\columnwidth]{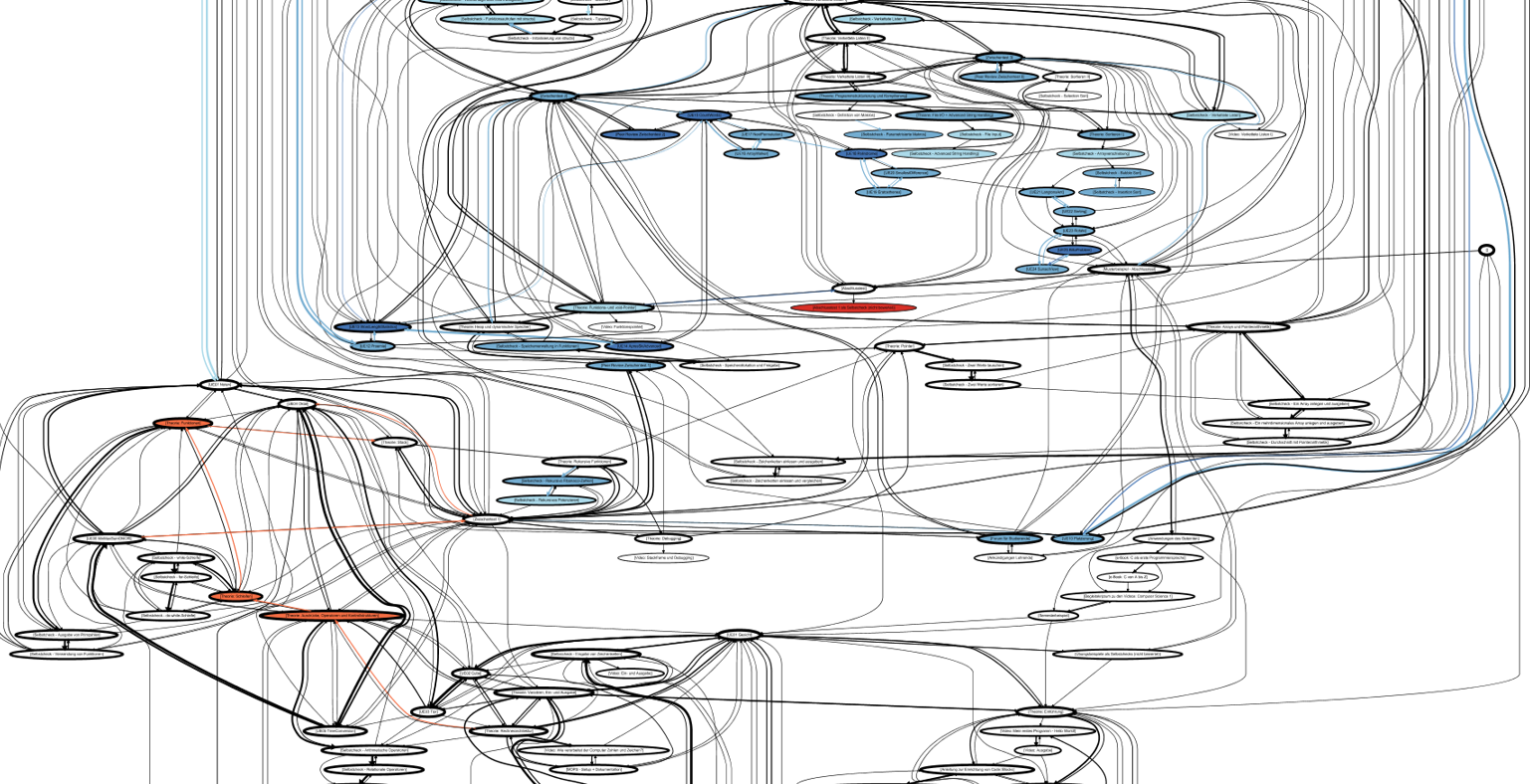}}
	\caption{Detail of Procedural Languages filtered.}
	\label{fig:proc2}
\end{figure}

Figure \ref{fig:proc1} illustrates the outcome of the Process Comparator using the settings described before, without filtering events or transitions. Due to the complexity of the visualization, the built-in filter of the plugin was subsequently adjusted to remove the least-frequent 10\% of activities and paths for all subsequent analyses. This refinement is depicted in Figure \ref{fig:proc2}.

Red nodes indicate events that occurred more frequently in \textbf{group B} (the lower-scoring half of the students), while blue nodes represent events that occurred more frequently in \textbf{group A}, the higher-scoring half of the students. Additionally, red edges denote paths taken more often in group B, whereas blue edges indicate paths taken more often in group A.
Five activities, including theory and self-checks, occurred more often in group B, whereas a significantly larger number of activities occurred more frequently in group A. Moreover, the majority of paths (edges) that occurred significantly more often in a specific group can be attributed to group A.

In both groups, the paths that occur more frequently in one of the two groups predominantly consist of routes leading to those activities, with very few exceptions observed (such as transitions from intermediate tests to theory or back-and-forth movements between different theory resources) in both groups.

\begin{figure}[htbp]
	\centerline{\includegraphics[width=\columnwidth]{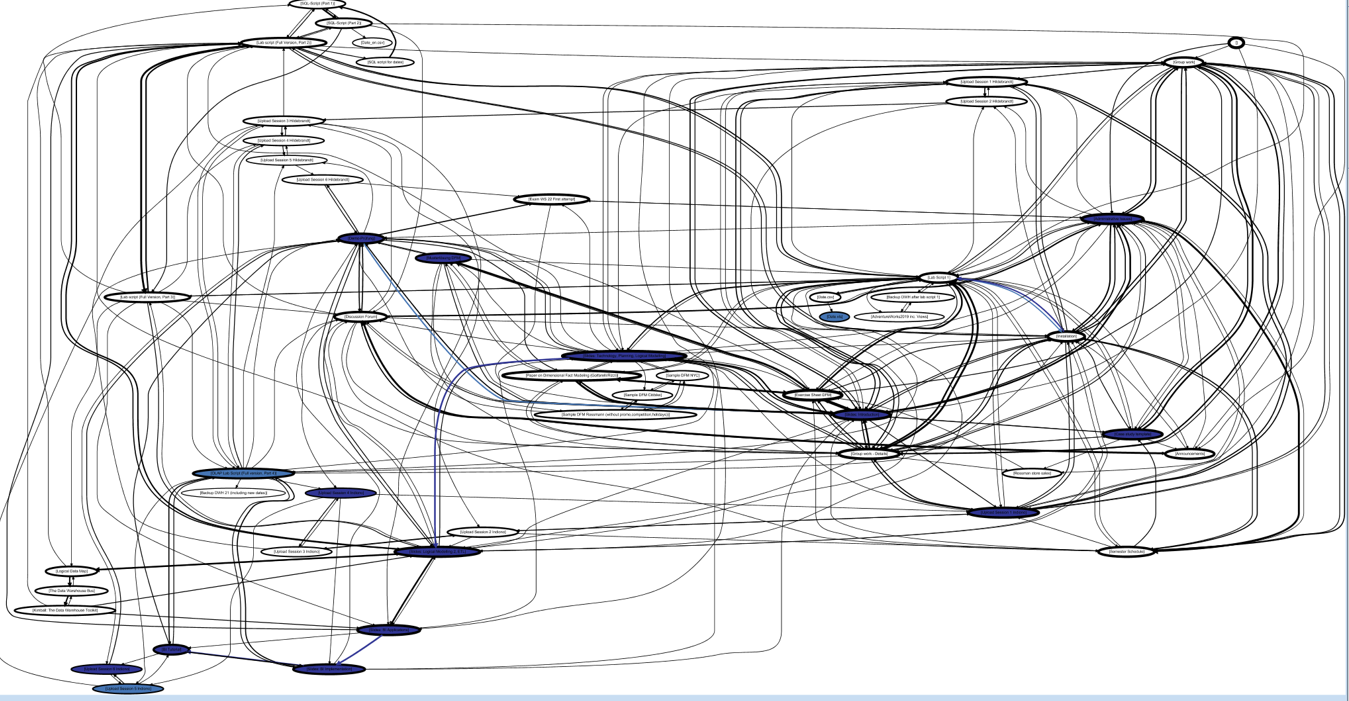}}
	\caption{Data Warehouse \& Business Intelligence path comparison.}
	\label{fig:dwbi}
\end{figure}

For the course \textbf{Data Warehouse \& Business Intelligence}, the cutoff grade was set at 79.2 out of 100 points, with a total of 5,874 events recorded. Group B accounted for 2,140 events, while the higher-performing group A generated 3,734 events. For reference, please see Figure \ref{fig:dwbi}.

The results were similar to those of the analysis of the course Procedural Languages: the higher-performing students of group A engaged in more optional activities and subsequently followed the paths leading to those activities more often as well. However, there were a few exceptions, such as paths leading back and forth between tutorials and installation manuals. Notably, there were no paths or activities that occurred more frequently in Group B.

In summary, an analysis solely based on individual course activities may not always be meaningful, as activities are typically grouped into sections within most courses, which are not considered in this analysis. Depending on the course, each section typically contains a handful of Moodle elements (the mean across all courses in our databases was 4.2 elements per course section) - some sections may contain only 1 element (such as an exam section), some sections can contain more than 100 elements. For course structure or achieving an ideal learning outcome, the order in which individual activities within a section are utilized may often be insignificant. Additionally, without grouping by sections, it becomes challenging to discern from a process map whether students are progressing "in the right direction" or moving "back and forth," as mentally mapping potentially hundreds of course activities into a sequence is complex. Therefore, it seems sensible to group activities by sections.

\subsubsection{Section-based aggregation}

\begin{figure}[htbp]
	\centerline{\includegraphics[width=\columnwidth]{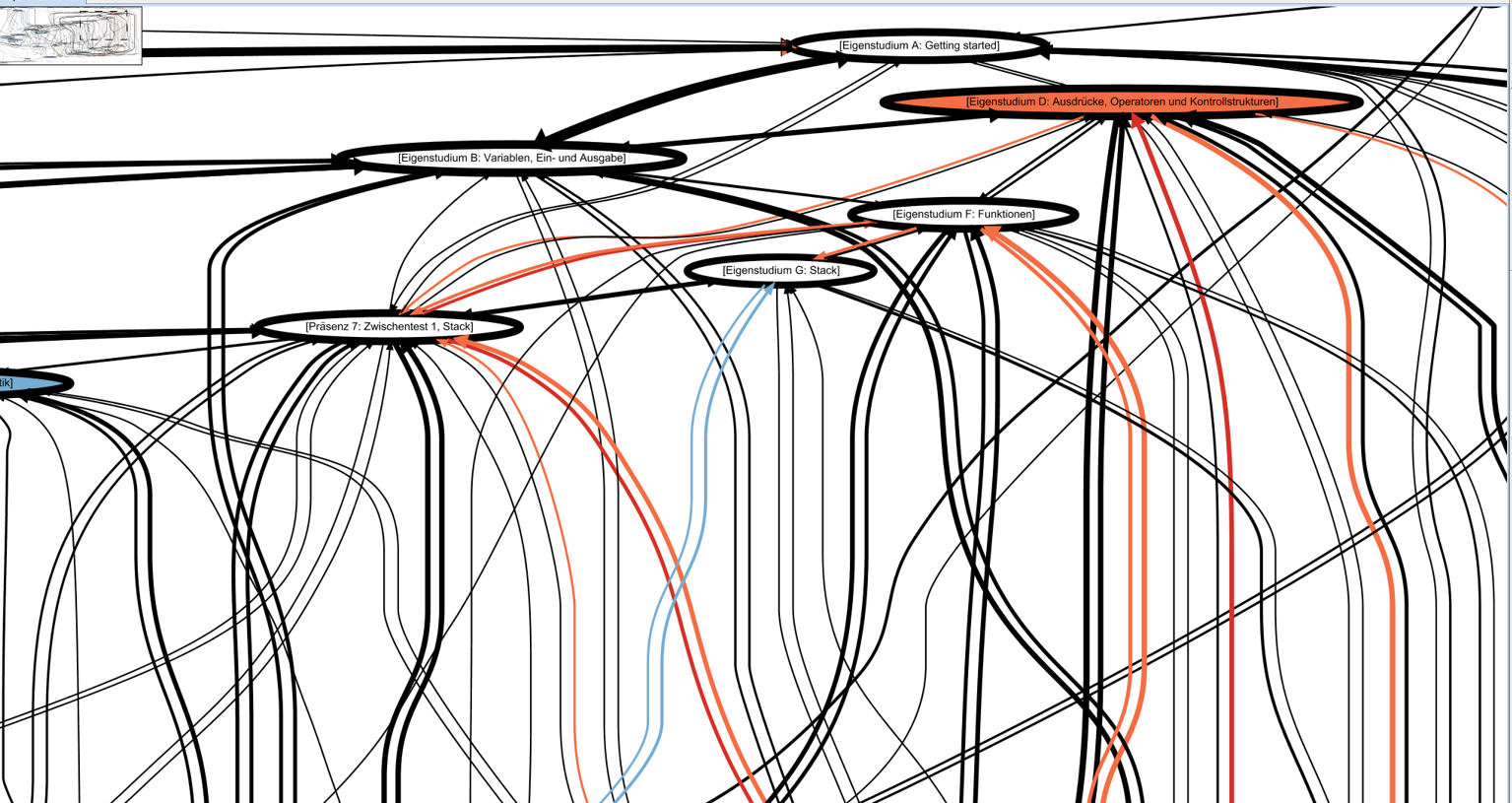}}
	\caption{Procedural Languages section-based}
	\label{fig:proc_section}
\end{figure}

When comparing sections instead of individual Moodle elements, similar to the previous analysis, for the course \textbf{Procedural Languages} there were more activities performed significantly more often by group A compared to group B (15 and 3 activities respectively). However, when examining the paths, the situation was less clear, appearing almost balanced. Paths taken more often by group A typically led to presence units or self-study sections. Conversely, paths taken more often by the lower-performing group often centered around an intermediate test, as depicted in Figure \ref{fig:proc_section}.

\begin{figure*}[htbp]
	\centerline{\includegraphics[width=0.9\textwidth]{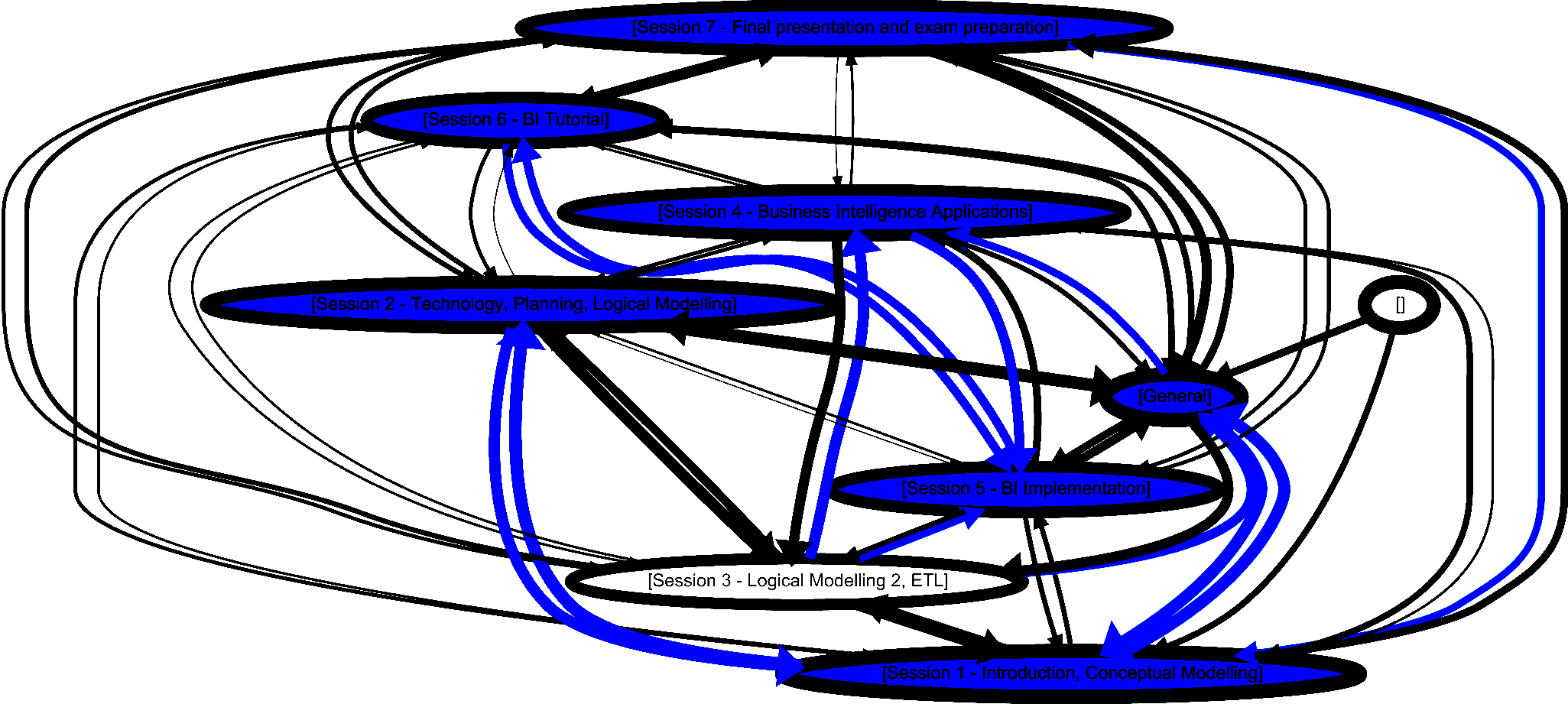}}
	\caption{Data Warehouse and BI section-based}
	\label{fig:dwbi_section}
\end{figure*}

Returning to the course \textbf{Data Warehouse and Business Intelligence} previously analyzed, the section-based analysis presents a much more organized process map, given that the course comprises only 8 sections. The results appear clearer compared to the analysis of individual elements, with all but one section being visited more frequently by group A (the higher-performing half). Additionally, students in group A demonstrate a higher frequency of movement back and forth between sections. Similar to the element-based analysis, the paths taken more often primarily lead to activities that are also clicked on more frequently, as depicted in Figure \ref{fig:dwbi_section}.

In general, similar patterns in terms of activity and path usage emerge when analyzing individual events compared to grouped sections. However, these patterns seem to be more clearly delineated in the section-based analysis. When analyzing individual elements, the need arises to filter out low-occurring events and paths to discern patterns in the process map. Conversely, during section-based analysis, filters are often unnecessary as the individual elements are already grouped and summarized under their respective sections.

\subsection{Course-spanning path comparison }
After the standardizing process describe earlier, 393 courses with 10,451 cases were included in the course-spanning analysis. In this instance, it was unnecessary to distinguish individual students, as some students were enrolled in multiple courses. Therefore, as described earlier, a unique identifier of the combination of courseID and studentID was created for each student in a course, resulting in a total of 881,441 events.

\begin{figure*}[htbp]
	\centerline{\includegraphics[width=\textwidth]{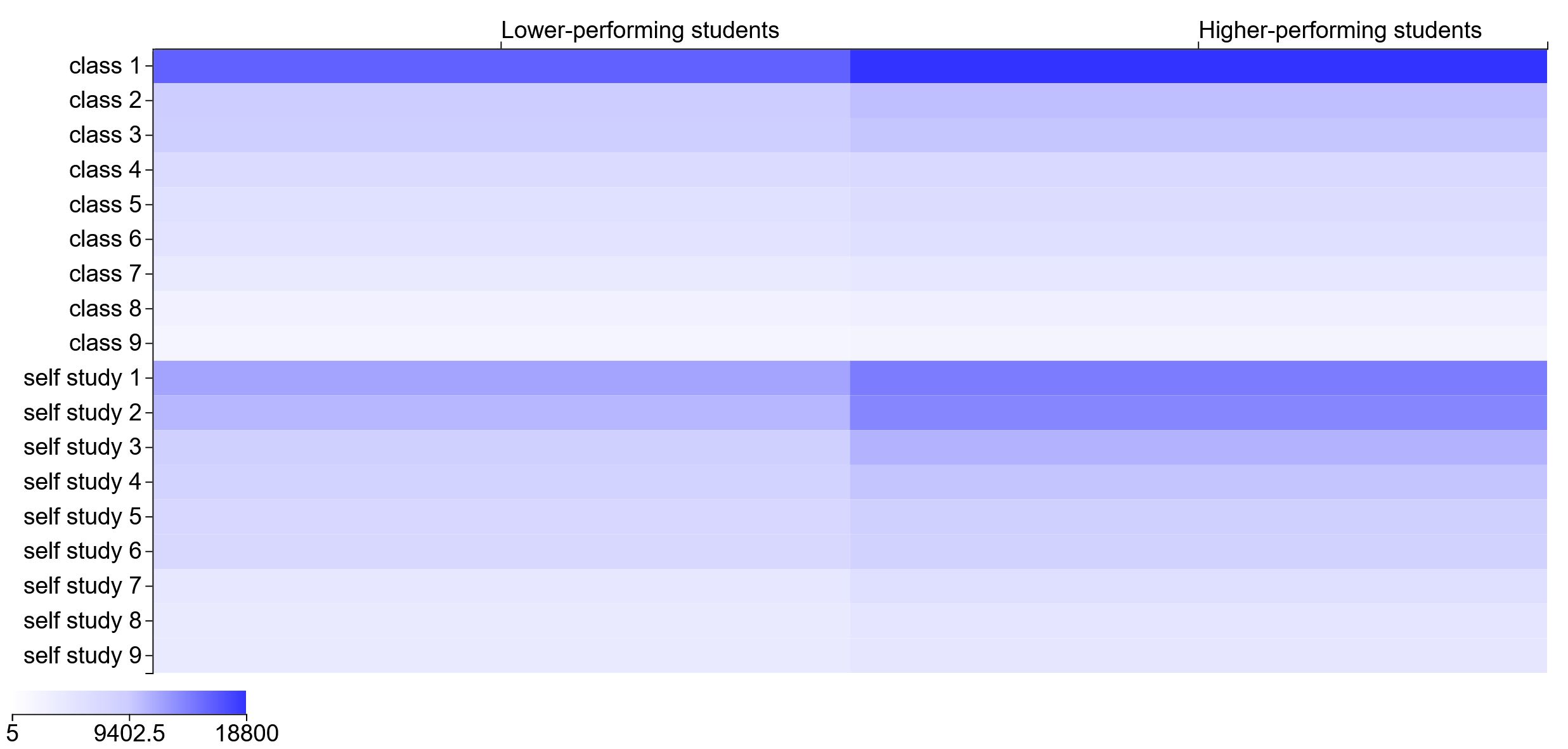}}
	\caption{Frequency of interactions across sections for both groups}
	\label{fig:heatmap}
\end{figure*}

The median score across the included courses was calculated at 79.33 out of 100. Group A (the better performing half of the students) comprised 454,065 events, and group B (the lower performing half of the students) comprised 346,376 events.
Across most activities, we observed a trend where better-performing students tended to visit them more frequently. This is summarized in Figure \ref{fig:heatmap}. In this heatmap, the differences concerning how often the members of both groups interacted with elements of the respective sections can be seen. What is furthermore apparent, is that for both groups there is a tendency of declining interaction with more advanced sections. 

\begin{figure*}[htbp]
	\centerline{\includegraphics[width=\textwidth]{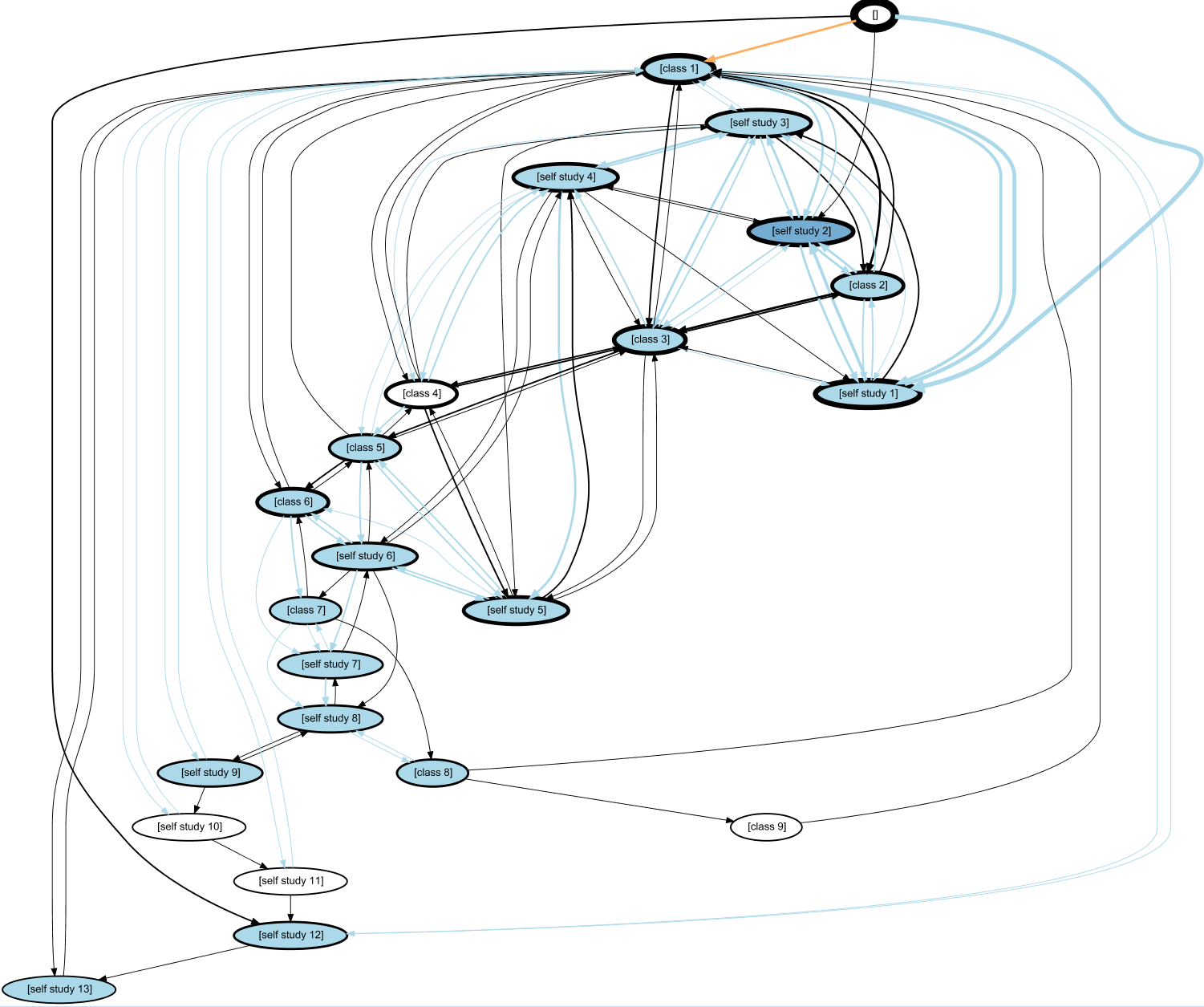}}
	\caption{Process Comparator course-spanning}
	\label{fig:course-spanning}
\end{figure*}

As can be seen in Figure \ref{fig:course-spanning}, in terms of paths traversed similar patterns as in the intra-course analytics can be observed. Better-performing students tended to explore more paths more frequently, with the exception of the path leading directly from the start of the course to class 1. These paths primarily consisted of those leading to the most frequently visited activities, such as transitions between classes and their corresponding self-study sessions, or progressions from one self-study session to the next. However, we noted a distinct pattern of back-and-forth movement between self-study sessions and the previous class, exemplified by transitions between class 2 and self-study 1. Furthermore, we noted that higher-achieving students exhibit a tendency to navigate more frequently between different sections, displaying a pattern of jumping back and forth more frequently than their lower-performing counterparts.

One conclusion that can be drawn is that it seems that better-performing students more often engage in self-study sessions prior to attending corresponding classes, whereas the students of group B more often skip the self-study sessions and go directly to the corresponding classes, a behavior that has been validated by analyzing the individual patterns in Fluxicon Disco.
As an example, the students of group A start the course with the section self study 1 in 64 \% of cases, while that is only the case in 32.1 \% of times in Group B. Overall, 86.1 \% of students of group A visit self study 1 at least once at some point in time (74.8 \% in Group B) and 78.3 \% of students of group A visit self study 2 at some point in time (59.2 \% in Group B).

\section{Discussion}
A substantial body of research leverages students' clickstream data extracted from learning management systems like Moodle. However, the majority of this research focuses on analyzing individual events within single courses. Only a few studies address the challenge of aggregating these individual events into higher-level entities, such as course sections. To the best of our knowledge, there is no research that deals with systematically standardizing activities across courses to facilitate cross-course analysis, such as comparing the learning paths of lower- and higher-performing students.

This paper introduces a toolchain and process to tackle these challenges. Our findings reveal that across various courses, higher-performing students tend to engage more extensively with course materials. Specifically, they demonstrate more frequent interactions with available activities such as resources, weblinks, upload fields, and forums compared to their lower-performing counterparts. This behavior becomes more apparent when individual events are grouped into course sections. Notably, we observed that higher-performing students exhibit a propensity to navigate more dynamically between different sections, indicating a more exploratory learning approach.

\begin{figure}[htbp]
	\centerline{\includegraphics[width=\columnwidth]{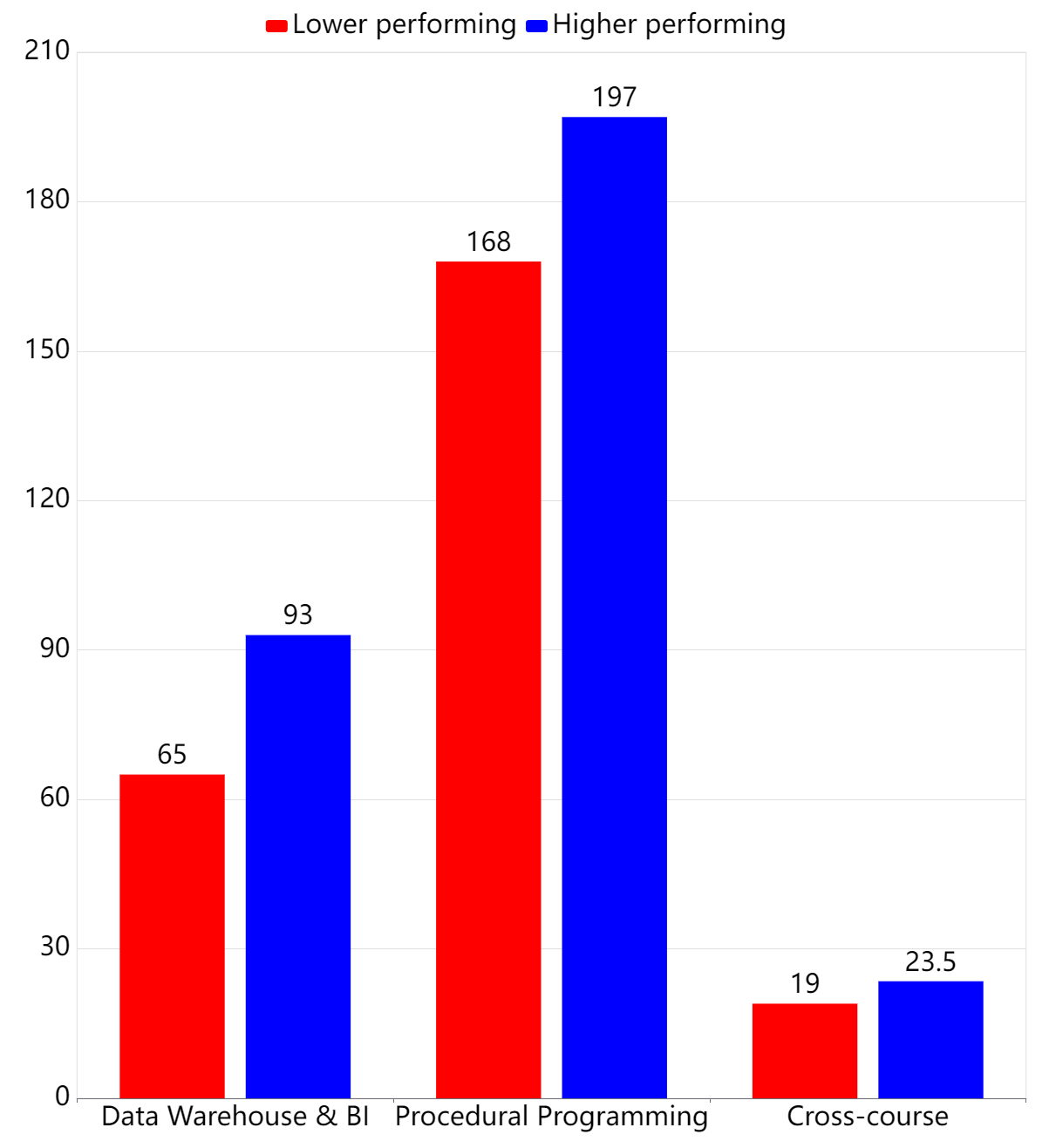}}
	\caption{Median number of section changes compared by groups}
	\label{fig:navigation}
\end{figure}

Moreover, this behavior extends to a course-spanning level, where higher-performing students demonstrate a pattern of frequent navigation between self-study units and corresponding course materials. 
Figure \ref{fig:navigation} summarizes this observation as it shows that members of the higher-performing groups navigate more often between course sections than members of the lower-performing half. In the cross-course analysis, the median number of section changes is roughly 23.7\% higher compared to the other group.

This finding underscores the importance of not only understanding students' interactions within individual courses but also analyzing their behavior across multiple courses to glean deeper insights into learning patterns and performance outcomes.
The causality behind this observed behavior was not within the scope of our research. It remains unclear whether higher-performing students excel due to inherent learning styles, pre-existing familiarity with learning materials, or simply because they engage more frequently with available resources. Nevertheless, our findings underscore the potential implications for further research. We discovered that patterns are sometimes more readily discernible when data is aggregated at the section level, suggesting that tools or plugins like the Process Comparator could should be enhanced to offer drill-down functionalities like they are available in Business Intelligence tools such as PowerBI or Tableau. Additionally, enhancing tool functionality to facilitate the exchange of activity columns within the interface would streamline the analytical process.

For practical application, one insight could be to optimize LMS course structures to facilitate seamless transitions between activities that experience significant traffic. This could involve integrating repetitions of crucial content into regular course structures or presence units, thereby encouraging students to revisit and reinforce their learning. The analysis results can help in restructuring the course to create a more logical and intuitive flow. For instance, if students often skip certain sections, it may indicate a need to reorganize the content or provide better transitions between topics.
Instructors can use this data to group related activities and materials, ensuring a coherent learning path that aligns with how students naturally progress through the course. 

Additionally, our cross-course analysis highlights the value of making it as effortless as possible for students to navigate between different sections. Strategies aimed at motivating and facilitating this behavior, such as integrating repetitions or encouraging exploration, could enhance learning outcomes across courses.

\section{Conclusion}
By analyzing LMS clickstream data with means of process mining, it can be seen that higher-performing students interact more with LMS courses than their lower-performing counterparts. Furthermore, better-performing students also move more back-and-forth between LMS sections. This behavior can be seen when performing an event-based analysis for individual courses, but even more so when performing the analysis on aggregated course sections (Research question II) or when performing course-spanning analysis (Research question III). 
By homogenizing section titles across courses (e.g. replacing the title of the first course section with \textit{Section 1}), it is possible to prepare the data for such course-spanning process analysis (Research question I).  
The question that still remains open, is if the relationship between LMS engagement and students performance is causal or merely a correlation. In any case, it seems sensible to design LMS courses in such a way as to make navigation between elements and sections as easy as possible.

%\section*{References}
\FloatBarrier
\bibliography{bib} % Entries are in the refs.bib file

\end{document}